# Dielectric permittivity of aqueous solutions of electrolytes probed by THz time-domain and FTIR spectroscopy

A. De Ninno[1*], E. Nikollari[2], M. Missori[3], F. Frezza[2]

[1] ENEA- Frascati Research Centre, via E. Fermi 45, Frascati 00044 (Roma) Italy

[2] Department of Information Engineering, Electronics, and Telecommunications, Sapienza University of Rome, Rome 00184, Italy

[3] Istituto dei Sistemi Complessi, Consiglio Nazionale delle Ricerche, Rome 00185, Italy

[*]Corresponding author

## Abstract

In spite of a large number of papers published in the last twenty years in the literature, the basic model of liquid water is still controversial. We have measured the dielectric permittivity of pure water and aqueous chlorides solutions in the range 0.2-1.5 THz and we have found that the overall dielectric behavior can be satisfactorily described with the weighted average of two single time-constant Debye functions. The weights of the two components have been obtained by infrared spectra analysis and are related to two populations of water molecules having a different long-range organization. The resulting functions allow to fit the experimental data for pure water and solutions of LiCl, KCl, NaCl, and CsCl and to predict the excess response on the high frequency side of the relaxation without "ad hoc" corrective terms. These findings point to the evidence of two fluids in liquid water at normal temperature and pressure.

1. Introduction

The structure of liquid water, hydration processes and ion-solvent interactions are of paramount importance in the understanding of solution chemistry, biochemical reactions and material science, in general. Understanding the molecular structure of liquid water and, in particular, the intermolecular hydrogen bond is a fundamental requirement for chemistry and biology but is also a powerful tool to explain the properties of the solutions and design meta-materials.

The molecular dynamics modelling of the shapes of liquid $H_2O$ seems to support the continuum model of water: liquid water structure is provided by a three-dimensional network of continuously evolving hydrogen bonds (HB) .[1,2] The concept of HB is merely phenomenological. The computed HB lifetime differs indeed depending on which event is considered as breaking it (whether a continuous or an intermittent criterion is assumed), leading to differences in the evaluation ranging (for a single HB) from 0.05 to 10 ps.[3] HB lifetime depends on which event is assumed to cause the breaking of the bond. If we consider the violation of the first H bond criterion (first molecules has hydrogen attached to its oxygen) as breaking this correspond to a continuous lifetime. However, the violation of criteria is not necessarily accompanied by HB breaking; it can be a consequence of the libration motion of two particles that remain connected by HBs as they move away from each other. The intermittent life-time criterion applies when the violation criteria is considered the breaking of HB followed by the formation of a new HB with a third particle. This makes a description of liquid water only in terms of HB quite questionable mostly in the absence of a shared HB definition. Many attempts to model hydrogen bonded networks have been done. However, the controversial definition of the HB lifetime makes it very hard to justify the lifetime of H-bonded water clusters detected despite the fact that experimental results do not support randomly oriented molecules or molecules with only short-range correlations [4,5]. In order to have picoseconds stable clusters of water molecules it is mandatory to envisage a picoseconds stable interaction mechanism, which involves many tens of molecules at one time. Furthermore, many experimental results point towards the existence of two different fluids in water normally identified as hydrogen bonded networks and individual molecular modes governed by collisional processes[6-9]. It has been suggested that the statement that HB bonding is the origin of the collective ground state of liquid water has in point of fact to be reversed. The existence of a collective ground state, different from the ground state of isolated molecules as predicted in the realm of quantum field theory is the physical cause of the phenomenological concept of HB[10].

Terahertz time-domain spectroscopy [11] has shown a sub-picosecond fast relaxation, in addition to the picoseconds relaxation attributed to the fluctuation of the HB networks. Molecular Dynamics simulations have assigned this fast relaxation to individual modes[12]. Data of the complex dielectric permittivity of water obtained at these frequencies are generally fitted with a Debye function with two relaxation times: a slow one and a fast one, associated to structured clusters and independent molecules respectively. In addition, in order to merge the two contributions, two relaxation strengths have been introduced for the two modes. However, at frequencies higher than 1 THz, as the data start to diverge from the predicted values, one can clearly see that the overall dielectric behavior cannot be reproduced by simply adding a second term to the Debye's formula. In order to fit the data it is mandatory to add a third relaxation mode and/or a stretching and libration modes. Thus, the permittivity of pure water is considered to result from the superposition of mainly four processes: slow relaxation, fast relaxation, intermolecular stretching vibration and intermolecular libration [13]. Still, the amount of the contributions of the vibrational and libration modes may be over-estimated because of the too many parameters in the fit. Furthermore, the effects of the two relaxation processes are accounted for in the fitting function by the addition of a second relaxation time to the Debye's formula and not as two self-standing contributions. As a result, the lower limit of the slow relaxation strength must coincide with the higher limit of the fast relaxation one. This apparently small detail makes conceptually a big difference in the interpretation of the dielectric behavior, and is the starting point of our approach to understanding the structure of liquid water.

In this paper, we focused on understanding the ions effect on the liquid water dynamics as it can be inferred from the THz time domain spectroscopy (THz-TDS). When ions dissolve in water, we expect the slow relaxation to be affected by the tight binding between the ion and water molecules in the first hydration shell. On the other hand, the fast relaxation is modified by the structure breaking effect increasing the fraction of the free water molecules according to information from infrared (IR) spectra [14] confirmed by THz [15] investigations. We measured the dielectric spectra of aqueous solutions of chlorides at different concentrations by means of THz-TDS in the range 0.2 to 1.6 THz. Section 2 describes the experimental setup for THz-TDS and summarizes the method used to obtain the permittivity of the solutions. Section 3 contains the analysis of the complex dielectric spectra of pure water and chloride solutions. In the concluding remarks, we discuss our results and the inhomogeneous structure of water at ambient conditions.

## 2. Materials and Methods

We have studied the transmission spectra in the range 0.2 - 1.6 THz of samples of pure water and of aqueous solutions of the following electrolytes: NaCl (0.6, 1, 2, 3, 4 and 5M), KCl (1, 2, 3 and 4M), CsCl (1, 2, 3 and 6M) and LiCl (1, 2, 3, 4, 7, 10 and 12M). All solutions were prepared using deionized water (2 μS conductivity) and the samples were measured in an air-conditioned room at the temperature of (25±0.5)°C and (44±2)% relative humidity.

THz spectra of all samples were obtained using a Menlo Systems TERA K15 spectrometer equipped with photoconductive antennas excited by a femtosecond fiber-coupled laser (Menlo Systems T-Light). The THz beam line of the system was designed by using a couple of identical HRFZ-Si plano-convex lenses with effective focal length f=25 mm. The first lens collects the diverging THz radiation emitted by the emitter antenna and collimates a 12 mm diameter beam on the sample. The second lens collects the radiation transmitted through the sample and focuses it on the detector antenna (fig.1). For all acquisitions, the scan range was set to 100 ps and the data were collected with a time resolution of Δt=33.3 fs.[16,17]

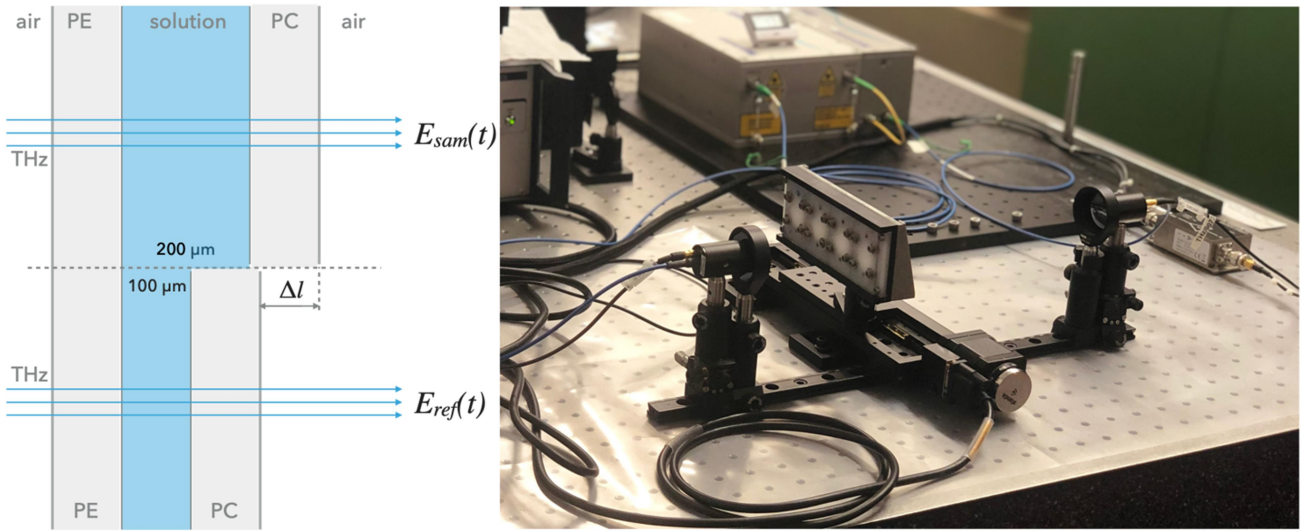

Fig. 1 Experimental setup. Left: permittivity data are obtained by comparing the signal collected through the sample cell with the one collected through the reference cell, the latter containing the same solution as the sample cell and differing from it only in the thickness of the solution. Right: The sample holder is placed in the path of a collimated THz beam. The temperature and relative humidity of the environment are kept constant at (25±0.5)°C and (44±2)% respectively.

The sample holder consists of four cells realized with two parallel slabs and a rhombic polyethylene spacer (100 or 200 μm thick) sandwiched between them, thus creating two cells for holding a 200 μm thick sample, and two more cells for holding a 100μm thick solution which is used as reference during the processing of the data. One of the two parallel slabs consists of a 5mm thick high density polyethylene (HDPE) and the other one of a 3mm thick polycarbonate (PC). While HDPE is quite transparent in the THz frequencies (indeed it is widely used as a window material in THz applications) and is opaque to the visible light, PC is not a good window material for the THz frequencies but it is transparent in the visible range. The compromise of using a transparent window material such as PC is dictated by the need to provide an economic mean for visually identifying and removing air bubbles contained in the liquid sample injected in the cells, and it is justified by the fact that the resulting S/N ratio does not affect significantly the quality of the measured data for the purposes of our study. The use of two different window materials instead, as well as the choice of not working in a water-vapor free environment, is justified by the differential approach used in analyzing the data as explained in the following, which minimizes the effects of common error sources and the lack of symmetry in the experimental design.

The dielectric parameters of the investigated solutions were obtained by comparing the signal collected through the sample (the 200 μm cell) with the one collected through the reference (the 100 μm cell filled with exactly the same material as the sample cell) (fig. 1). In such differential approach, when comparing the sample signal with the reference one by considering the ratio of their Fourier transforms, all common terms of the theoretical expressions of the sample and reference signals (the amplitude of the applied field $E_0$ and the transmission coefficients $\tau_{ij}$ at the interfaces between different materials) cancel out. This condition holds as long as the sample and reference cells are made of the same materials and both are measured in the same environmental conditions:

$$\frac{E_{sam}(\omega)}{E_{ref}(\omega)} = \frac{E_0 t_{ap} t_{pw} t_{wp} t_{pa}}{E_0 t_{ap} t_{pw} t_{wp} t_{pa}} e^{-i\omega\Delta\varphi^*} = e^{-i\frac{\omega\Delta l}{c}(n^*-1)} \xrightarrow{yiel} \frac{A_{sam} e^{i\varphi_{sam}}}{A_{ref} e^{i\varphi_{ref}}} = X e^{i\Theta} \quad (1)$$

where $\omega = 2\pi f$, $\Delta l$ is the path-length difference between the sample and reference signals, $n^* = n' - in''$ is the complex index of refraction of the solution, $X = A_{sam}/A_{ref}$ is the ratio of the amplitudes of the two Fourier-transformed signals and $\Theta = \varphi_{sam} - \varphi_{ref}$ is their phase difference.

This produces a straightforward mathematical expression for the complex index of refraction of the investigated solution without needing to characterize the window materials: the real part $n'$ results to be a simple linear function of only the phase difference $\Theta$ and the imaginary part $n''$ is a logarithmic function of only the ratio of amplitudes X:

$$\frac{\omega\Delta l}{c}(n'-1) = \Theta \xrightarrow{yields} n' = 1 - \frac{c}{\omega\Delta l}\Theta \quad (1a)$$

$$e^{-i\omega\frac{\Delta l}{c}n''} = X \xrightarrow{yields} n'' = -\frac{c}{\omega\Delta l}\ln(X) \quad (1b)$$

where $c$ is the speed of light in vacuum.

The permittivity is then obtained as:

$$\varepsilon^*(\omega) = n^{*2}(\omega) = \left(n'^2(\omega) - n''^2(\omega)\right) - i2n'^2(\omega)n''^2(\omega) \qquad (2)$$

Each measurement session consisted in filling the four cells with the same solution (resulting in two sample-reference couples) and in measuring each cell in sequence during the same session over an integrating period of 100 seconds per cell, for an improved S/N ratio. All collected signals were then windowed over a 20 ps interval centered around the peak of the THz pulse using a Hanning function before being Fourier transformed. This procedure was repeated 3 times for each solution resulting in 6 measurements per sample whose average produced the final values of the complex permittivity of the solution.

## 3. Results and Discussion

The exponentially decaying autocorrelation function of the polarization results in the generalized Debye-type relaxation spectral function in the frequency domain:

$$\varepsilon(\omega) = \varepsilon_\infty + \sum_{j=1}^{n} \frac{(\varepsilon_0 - \varepsilon_j)}{1 + i\omega\tau_j} \qquad (3)$$

In equation (3) $n$ separable processes j contribute with relaxation times $\tau_j$ and relaxation strength $(\varepsilon_0 - \varepsilon_j)$ [18].

According to Debye's model, the polarization developed by a system of non-interacting dipoles in response to the influence of an external field relaxes exponentially as a consequence of thermal fluctuations once the field is switched off. This equation describes the relaxation of $n$ sets of equivalent, non-interacting dipoles characterized by a single relaxation time $\tau_j$ from the measured static dielectric constant $\varepsilon_0$ to the residual independent dipoles permittivity $\varepsilon_\infty$ at very high frequencies. In dielectric relaxation studies appeared in the literature, a good agreement has been found between experimental data and the Debye model, by assuming a single exponential relaxation of the polarization for frequencies up to 100 GHz. If, however, the Debye model is extrapolated to THz frequencies, it fails to reproduce the experimental findings. In an attempt to account for the response of water at THz frequencies, a double Debye model ($n = 2$) has been adopted with four temperature-depending parameters: two relaxation times $\tau_1$ and $\tau_2$ and two relaxation strength $(\varepsilon_0 - \varepsilon_1)$ and $(\varepsilon_1 - \varepsilon_\infty)$.

The slowest process has been generally assigned to arise from the collective structure of liquid, i.e. from the response of the water molecule involved in a tetrahedral arrangement. The fastest has been assigned to the rotation of a single (not hydrogen-bonded) water molecule [19,20,21]. In such a picture, the residual collective polarization coincides with the low-frequency response of non-interacting dipoles and is represented by the parameter $\varepsilon_1$ which appears as an intermediate step in the relaxation from $\varepsilon_0$ to $\varepsilon_\infty$

More generally, it is widely accepted that dielectric materials have two different microscopic structures: one composed of atoms or molecules bonded by normal chemical bonds and another partially ordered liquid-like phase. When placed in an electric field the two different structures produce two separate responses [22].

The multiple relaxation model provides a good agreement with the experimental data up to 1 THz despite a certain discrepancy in literature among the evaluation of the second relaxation time $\tau_2$ (ranging between 0.19[23] and 1.20[24] ps) which has been attributed to different experimental spectral ranges. Actually, the fast limit of the time scale for a single water molecule reorientation (free rotator) is expected to be ≈ 0.1ps. The measured relaxation time of long-range structured clusters exceeds the single dipole by a factor ranging between 8 (Barthel) and 50 (Rønne) as obtained by fitting the experimental data with a Debye function with two or three time constants. It has been also suggested that the slow process involves structural rearrangement and subsequent reorientation. At frequencies higher than 1 THz, however, the above model is not sufficient to fit the experimental curves and several vibration related processes [13, 25] and phononic or intramolecular resonances [26] have been evoked in order to fit the data. A master curve able to fit the data over a suitable temperature range has also been proposed in order to improve the understanding of the dielectric behavior in solids and liquids but it results in an empirical distribution of three parameters to be used in the generalized theory[27].

Considering liquid water as a two components fluid with different characteristics in terms of local structures we propose to abandon the generally used Debye model of eq. 1, which consists of a Debye function modified by the addition of one or more time constants, and fit the data with a superposition of two separate, single-time-constant exponential relaxations. Each component of liquid water contributes to the total polarization at an extent determined by which fraction of the liquid it constitutes. Thus, the resulting fitting function will be the weighed sum of two separate Debye functions, the weights being obtained by the interpretation of IR spectra of liquid water as explained below. The main feature of this approach is that it depicts the observed complex permittivity as the weighted average of 2 underlying processes relative to two separate phases. As a first consequence, we cannot set $\varepsilon_0 = 80$ because the experimental value emerges as an average between the static permittivity of the two components.

By combining dielectric, microwave, THz, and far-infrared spectroscopy we aim to get a broadband description of the dielectric behavior of liquid water. The analysis of ATR-FTIR spectra of liquid water shows a broad absorbing band in the range 2800-3800 cm$^{-1}$ assigned to the OH stretching mode of molecules involved in different hydrogen bonds. Low-energy vibrations are related to molecules forming larger clusters or stronger H-bonds, and on the opposite, high energy vibrations have been assigned to water molecules not involved in water-water correlation. Larger clusters, ice-like components, have been located at about 3200 cm$^{-1}$, smaller clusters (dimers or trimers) at about 3400 cm$^{-1}$, and monomeric water molecules, gas-like components, at about 3600 cm$^{-1}$ [28]. It can be also shown that a unique de-convolution of the band is possible with a minimum number of sub-bands centered respectively at (about) 3200, 3400 and 3600 cm$^{-1}$ by both using the second derivative of the spectrum[29] and applying the Multivariate Curve Resolution (MCR). We use the normalized areas identified by each de-convolution curve to obtain the relative fraction of each population in liquid water. The normalized area of the lowest energy sub-band, N is proportional to the ratio of highly correlated molecules versus the total number of molecules, while (1-N) is assumed to be proportional to the loosely correlated or uncorrelated fraction [30] (see details in the Appendix A). In Table I are shown the N values for pure water and chlorides at different concentrations. As it can be seen in Fig. 2 N depends on both the type and the concentration of the solute.

We can thus write down the equation for the relaxation spectral function:

$$\varepsilon(\omega) = N\left[\varepsilon_{\infty 1} + \frac{(\varepsilon_{01}-\varepsilon_{\infty 1})}{1+i\omega\tau_1}\right] + (1-N)\left[\varepsilon_{\infty 2} + \frac{(\varepsilon_{02}-\varepsilon_{\infty 2})}{1+i\omega\tau_2}\right] \quad (4)$$

The first term describes the Debye relaxation of clusters of highly correlated molecules while the second one describes the behavior of an assembly of independent dipoles distributed about the applied field in accordance with Boltzmann's law. It is straightforward to separate the real and imaginary part:

$$\varepsilon(\omega)' = N\left(\varepsilon_{\infty 1} + \frac{\varepsilon_{01}-\varepsilon_{\infty 1}}{1+(\omega\tau_1)^2}\right) + (1-N)\left(\varepsilon_{\infty 2} + \frac{\varepsilon_{02}-\varepsilon_{\infty 2}}{1+(\omega\tau_2)^2}\right) \quad (4a)$$

and

$$\varepsilon(\omega)'' = N\left(\frac{\varepsilon_{01}-\varepsilon_{\infty 1}}{1+(\omega\tau_1)^2} \cdot \omega\tau\right) + (1-N)\left(\frac{\varepsilon_{02}-\varepsilon_{\infty 2}}{1+(\omega\tau_1)^2} \cdot \omega\tau\right) \quad (4b)$$

Table I. Values of the normalized number of highly correlated molecules N of chlorides solutions.

| M | LiCl N | NaCl N | KCl N | CsCl N |
|---|---|---|---|---|
| 0.00 | 0.484 | 0.484 | 0.490 | 0.484 |
| 0.10 | 0.479 | 0.478 | 0.485 | 0.478 |
| 0.20 | 0.478 | 0.479 |  | 0.474 |
| 0.50 | 0.471 | 0.468 | 0.475 | 0.481 |
| 1.00 | 0.466 | 0.454 | 0.460 | 0.461 |
| 2.00 | 0.454 | 0.422 | 0.435 | 0.432 |
| 3.00 | 0.440 | 0.391 | 0.394 | 0.409 |
| 4.00 | 0.423 | 0.363 |  | 0.380 |
| 6.00 | 0.404 |  |  | 0.335 |
| 10.00 | 0.371 |  |  |  |

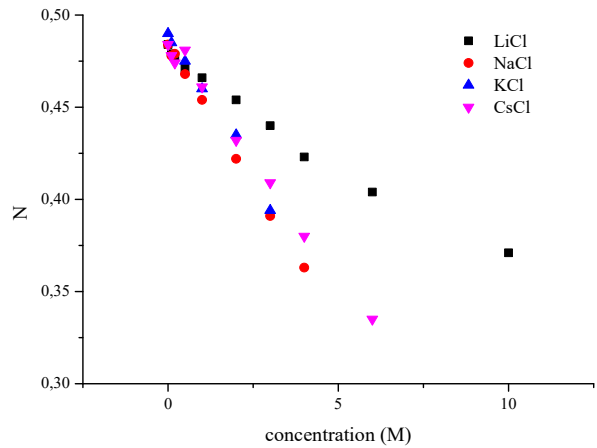

**Figure 2.** Plot of the N values versus the molar concentration of the chlorides solutions.

a. _Permittivity of pure water_

In order to fit the whole range relaxation spectrum of pure water we combined our data in the THz region with the data obtained by Kaatze in 1989[31] and used eq. (4a) and (4b) with N=0.484 as obtained by FTIR spectroscopy (see Table I) thus obtaining $\varepsilon_{01}$, $\tau_1$; $\varepsilon_{02}$; $\tau_2$, $\varepsilon_{\infty 1}$ and $\varepsilon_{\infty 2}$.

From this first step of data analysis we observed that $\varepsilon_{01}/\varepsilon_{02} \approx \tau_1/\tau_2$. We thus assumed that an ideal experiment would have yielded $\varepsilon_{01}/\varepsilon_{02} = \tau_1/\tau_2 = R$, where $R$ can be considered a scale factor between the dynamics of the two fractions, being $\tau_1 = R\tau_2$, and $\varepsilon_{01} = R\varepsilon_{02}$ i.e. we make the hypothesis that the same scaled phenomenon applies to two different populations having different characteristics in terms of local structures. The scale factor R cannot be attributed to the size of the cluster uless the principle of superposition strictly holds. R would actually be the number of molecules in a cluster if and only if collective effects are not taken into account. Without a satisfactory description of the long-range structure of liquid water, we cannot attribute to R any clear physical meaning. Rescaling the eq. (4) we obtain:

$$\varepsilon(\omega) = N\left[\varepsilon_{\infty 1} + \frac{(R\varepsilon_{02} - \varepsilon_{\infty 1})}{1 + i\omega R\tau_2}\right] + (1-N)\left[\varepsilon_{\infty 2} + \frac{(\varepsilon_{02} - \varepsilon_{\infty 2})}{1 + i\omega \tau_2}\right] \quad (5)$$

and

$$\varepsilon(\omega)' = N\left[\varepsilon_{\infty 1} + \frac{(R\varepsilon_{02} - \varepsilon_{\infty 1})}{1 + (\omega R\tau_2)^2}\right] + (1-N)\left[\varepsilon_{\infty 2} + \frac{(\varepsilon_{02} - \varepsilon_{\infty 2})}{1 + (\omega \tau_2)^2}\right] \quad (5a)$$

Fitting the data with this equation yields $\varepsilon_{\infty 2} = 1.9 \pm 0.1$, in contrast with the results of the fits reported in literature where a Debye fuction with an added time constant is used yielding for $\varepsilon_\infty$ values between 4 and 5.[11] Our result is consistent with the observation that, since the distortion polarization is established very quickly and represents the residual polarization at very high frequencies, $\varepsilon_{\infty 2} \sim n^2$ being $n$ the refraction index. We thus assumed in the fit the residual polarization of the un-correlated fraction to be coincident with $n^2$. In the case of water, at room temperature $n^2 = 1.78$. The second step consisted in a further least-squared fit of the (5a) with $\varepsilon_{\infty 2} = n^2$. (results are shown in Table II)

The fit of the imaginary part deserves a closer attention. In fact, it is not straightforward to obtain a good fit for the imaginary part even using the same parameters obtained by the least-squared fit of the real part. Such a problem has also been faced by other authors: Shcherbakov et al. observed that the relative error in the evaluation of $\varepsilon''$ is higher that of $\varepsilon'$ and increases with increasing frequency.[32] We find that a good match of the experimental data may be obtained both for the real and the imaginary part but the parameters of the fit significantly differ between the two or, vice versa, forcing the parameters obtained with the real part the imaginary part cannot be fitted. In order to face this problem we added a linear term $\zeta\omega$ to $\varepsilon''(\omega)$:

$$\varepsilon(\omega)'' = N\left[\frac{(R\varepsilon_{01} - \varepsilon_{\infty 1})}{1 + (\omega R\tau_1)^2} \cdot \omega R\tau_1\right] + (1-N)\left[\frac{(\varepsilon_{02} - \varepsilon_{\infty 2})}{1 + (\omega \tau_2)^2} \cdot \omega \tau_2\right] + \zeta\omega \quad (5b)$$

It must be underlined that the addition of the term $\zeta\omega$ appears like a violation of the Kramers-Kronig relation for a time span $\xi$. The Kramers-Kronig relation expresses the real part of the amplitude for forward scattering by molecules of light of a fixed frequency ω as an integral over the cross section for absorption by molecules of light of all frequencies. It has been derived by requiring the temporally non local connection between the displacement **D(x,t)** and the electric field **E(x,t)**:

$$D(x,t) = E(x,t) + \int_0^\infty G(\tau)E(x, t - \tau)d\tau$$

and

$$\varepsilon(\omega) = 1 + \int_0^\infty G(\tau)e^{i\omega\tau}\,d\tau$$

This shows that $\varepsilon(\omega)$ is an analytical function of ω in the upper half plane, provided that $G(\tau)$ is finite for all $\tau$. This is true for dielectrics but not for conductors, where $G(\tau) \to 4\pi\sigma$ as $\tau \to \infty$ and $\varepsilon \to i4\pi\sigma/\omega$ because of the contribution of the fraction of "free" electrons per molecules. In Table II are shown the parameters obtained by the fit with 5a and 5b. By using these parameters we obtained $\zeta = 0.35\ ps$ and a very good match of the fit with the experimental data.

In Fig. 3 is shown the complex dielectric relaxation of water at room temperature, $\varepsilon'(\omega)$ and $\varepsilon''(\omega)$, as a superposition of two weighted processes represented by the solid colored lines. The dashed coloured lines represent the maximum possible contribution of each process (for N=1 in the case of the slower fraction and for N=0 in the case of the faster one). The solid black line represents the fit function from eq. 5a and 5b and the dashed black line represents the linear term $\zeta\omega$.

We can observe that the values of $\tau_1 = 8.27\ ps$ and $\tau_2 = 0.31\ ps$ are consistent with the most recent studies [33,34].

TABLE II. Parameters used in the fit of eq. 5a and 5b for pure water. Constrains: N=0.484; $\varepsilon_{\infty 2}$=1.78; $\varepsilon_{01}/\varepsilon_{02} = R$

|  | $\varepsilon'$ | $\varepsilon''$ |
|---|---|---|
| $\varepsilon_{01}$ | 155.8 ± 0.2 | 156.7 ± 0.1 |
| $\varepsilon_{02}$ | 5.8 ± 0.2 | 6.7 ± 0.2 |
| $\varepsilon_{\infty 1}$ | 6.21± 0.04 | 6.21 ± 0 |
| $\varepsilon_{\infty 2}$ | 1.78 ± 0 | 1.78 ± 0 |
| $\tau_1$ | 8.27 ± 0.02 ps | 8.2 ± 0.3 ps |
| $\tau_2$ | 0.31± 0.01 ps | 0.31 ± 0.01 ps |
| N | 0.48 ± 0.01 | 0.48 ± 0.01 |
| R | 26.7 ± 0.8 | 23.3 ± 0.3 |
| $\xi$ |  | 0.47 ± ps |

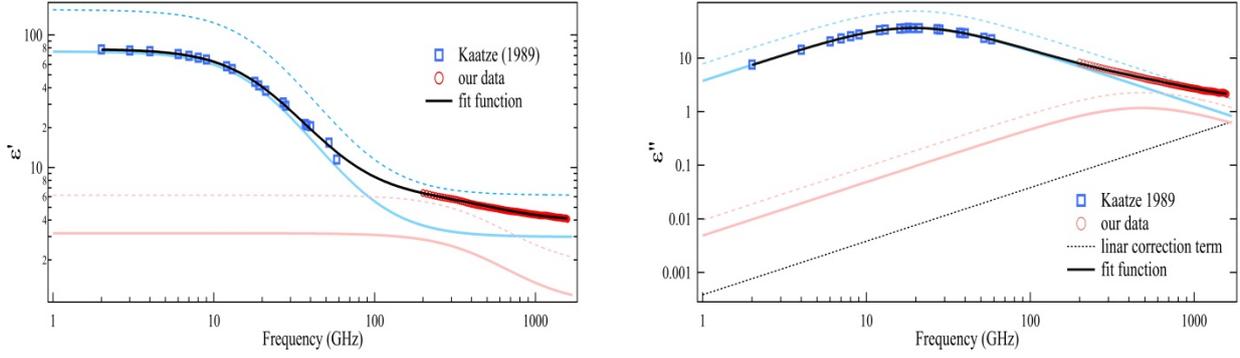

**Figure 3** Real ε' and imaginary ε'' parts of the dielectric permittivity of pure water as a superposition of two Debye processes, a slow one weighed by the factor N (solid light blue line), and a faster one weighed by the factor (1-N) (solid pink line). The dashed coloured lines represent the maximum possible contribution of the respective processes (for N=1 in the case of the slower fraction and for N=0 in the case of the faster one). The solid black lines represent the fit functions from eq. 5a and 5b and the dashed black line represents the linear term necessarily added in equation 5b in in order to fit the data

b. <u>Permittivity of chlorides solutions</u>

It is known that the addition of solutes affects local structures of water according to the concentration so, like in the case of pure water we have assumed $\tau_1 = R\tau_2$ and $\varepsilon_{01} = R\varepsilon_{02}$ expecting the calculated R to be related to the concentration . In order to confirm the validity of the procedure, we have applied eq. (5) also to aqueous solutions of chlorides at different concentrations. However, for the electrolytic solutions we could not find in literature sufficient low frequency experimental data in order to directly fit the permittivity over a wide frequency range as in the case of pure water. We have thus fit the data with the equation (5) setting $\varepsilon_{01} = \frac{R}{N(R-1)+1}\varepsilon_{0meas}$ and $\varepsilon_{02} = \frac{1}{N(R-1)+1}\varepsilon_{0mea}$ (as obtained from equation (5) in the limit $\omega \to 0$) where $\varepsilon_{0meas}$ is the measured static permittivity of the solutions, whose values are available in the literature. Then we weight the two components with the values N and (1-N) of Table I. It is worth noting that this approach allows to get rid of the high variability of the parameters and of the need to fix the value of $\tau_1$ choosing among a wide range of values reported in literature, which would strongly influence the results of the fit. Like in the case of pure water, we also set $\varepsilon_{\infty 2} = n^2$, being the refractive index of the solutions available in the literature.

From a first fitting cycle we observed that $\tau_2$ ranges between 22 and 46 fs in a random fashion, which suggested the need to introduce one further constraint. Since $\tau_2$ is the relaxation time of a free dipole it does not carry information about solvation shells or long-range restructuring of water induced by the solute. It should not be significantly different from the $\tau_2$ of pure water ($\tau_2 = 31\ fs$). We thus made the hypothesis that in all the solutions $\tau_2$ must coincide with the value measured for pure water. On the opposite, $\tau_1$ holds information about the collective behavior of water molecules facing the solvation of solutes.

The generalized permittivity for electrolytes is commonly represented with the help of the frequency-dependent complex quantity $\eta^*(\omega)= \varepsilon^*(\omega)+\frac{\sigma}{i\omega\varepsilon_0}$, where σ is the static conductivity of the solution, so we include the term $\frac{\sigma}{i\omega\varepsilon_0}$ in equation (5b) in order to account for the ionic contribution to the dielectric loss and set σ to values available in literature.

It is worthwhile to note that the values obtained for $\tau_1$ in NaCl are consistent with the findings of Shcherbakov[32] and quite lower than other results reported in literature[35,36]. For the other chlorides too, we predict values of $\tau_1$ which are lower than the ones reported in literature[36,37,38]
In Fig. 4 are shown the real and imaginary permittivity of chlorides. The match of the experimental and calculated values is very good and supports the claims introduced in the discussion in order to achieve a general description of the dielectric properties of solutions. In Fig.5 is shown the comparison among the permittivity of different salts at 3M. It is very interesting to underline the strange behavior of LiCl: it is the only salt to show a dissipation (see the $\varepsilon''$ panel) lower than pure water, in other words, it is more transparent than water at the investigated frequencies and below 1THz. Jepsen and Merbold [36] discussed the main differences between the behavior of the dielectric function of LiCl solutions and the other chlorides: a decrease of $\varepsilon''$ with concentration, whereas an increase for the other salts is observed. This is due to to the behavior of slow and fast relaxation processes: in Fig. 2 it can be seen that the LiCl influence on clustered water deviates from the other chlorides. The, so called, slow relaxation fraction decreases slower than the other salts with the concentration and this fact, obviously, affects the strength of the relaxation processes, as suggested. However, the reason has to be found in the different long-range restructuring of water induced by the Li ion because of its very high charge density.
The parameters of the fit for all the ions are shown in the Appendix B.

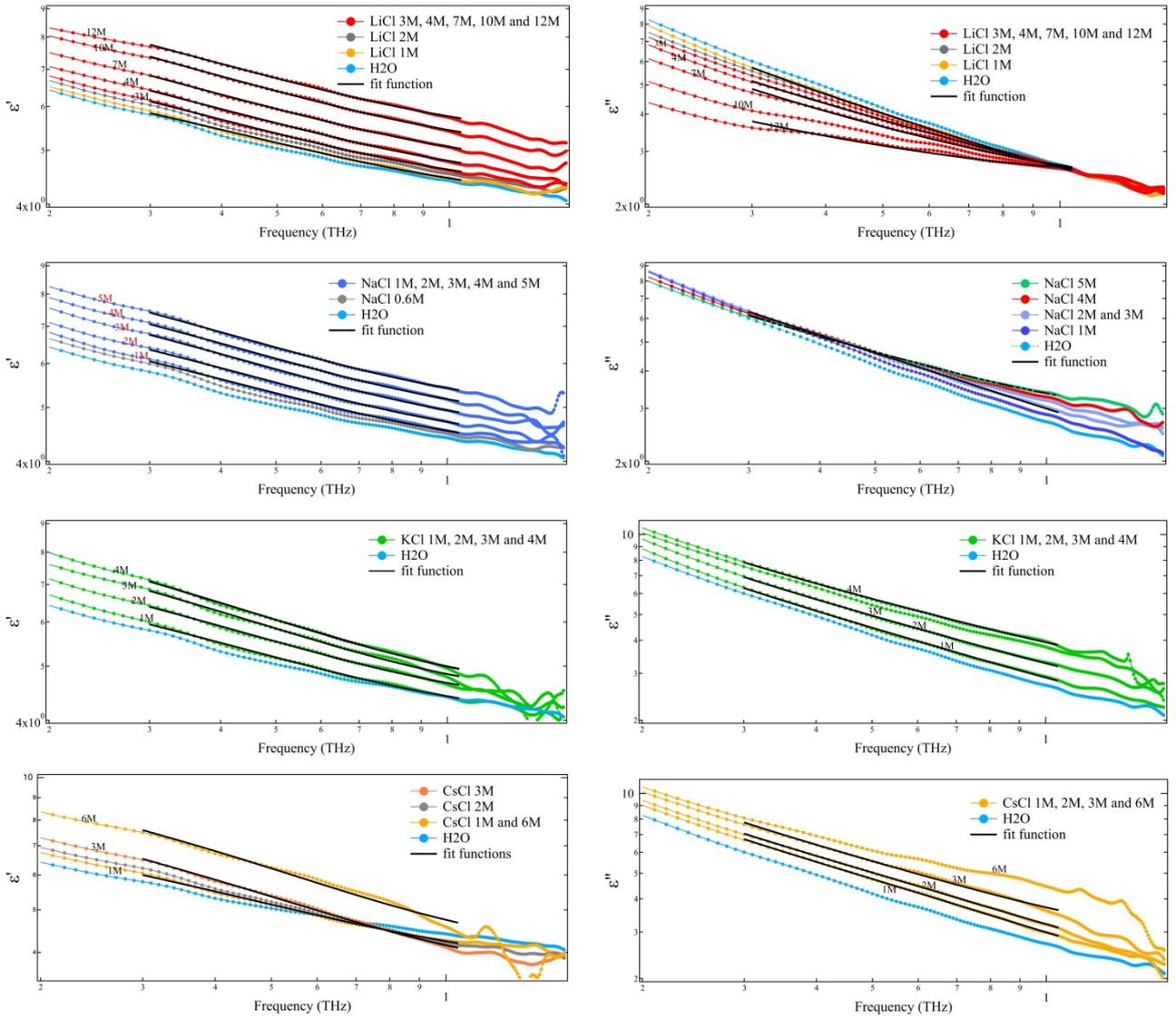

**Figure 4**. Real ε' and imaginary ε'' parts of the dielectric permittivity of each chloride solution at room temperature. The experimental spectra are shown by dots. Solid lines represents the fit functions from eq. 3a and 3b. The fit functions of the loss spectra have been obtained by subtracting $\frac{\sigma}{\omega\varepsilon_0}$ from eq. 3b. The parameters of the fits are shown in the tables of Appendix B.

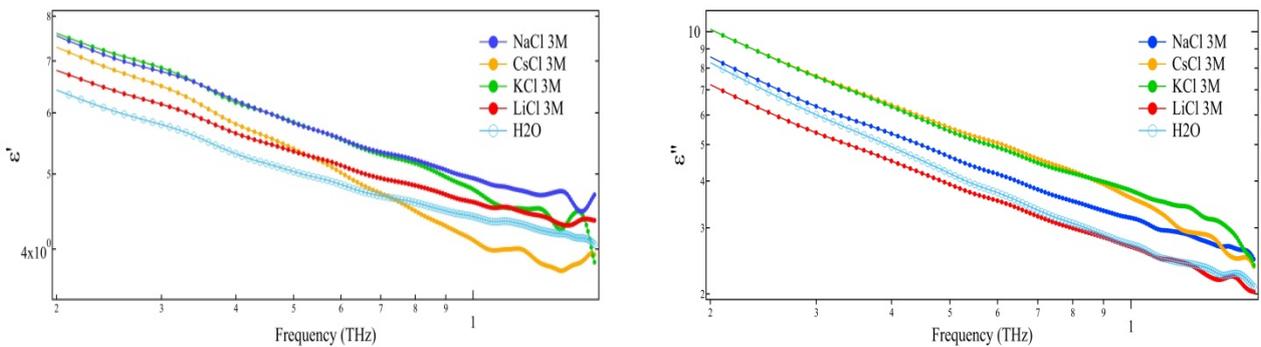

**Figure 5**. Comparison of the experimental data of dielectric permittivity of 3M chlorides.

## 4. Conclusions

We have measured and analyzed the dielectric response of pure water and chlorides solutions in the range 0.2-1.6 THz. The permittivity has been satisfactorily represented as the weighted sum of two single-parameter Debye relaxations where the weights are the relative abundance of two populations in liquid water. This approach also solved the problem of the excess high-frequency permittivity avoiding *ad hoc* terms added to the Debye function in order to fit the experimental data in the THz region[39]. The parameters obtained by the fitting describe clusters of correlated water molecules coexisting with poorly or uncorrelated single dipoles. The ratio R between the relaxation times is coincident with the ratio between the low frequency permittivity of the two fraction and provides the gauge of the clusters. We have found that R ≈ 26 for pure water and decreases almost linearly with the concentration of the solutes. This shows that the dielectric relaxation of a cluster of pure water molecules is 26 times slower with respect to the relaxation of a single poorly correlated molecule (monomer or dimer) and its polarization for orientation is very much higher, the experimental dielectric constant being just the average of two polarizations. This fact may be useful in the study of biological systems.

It is widely accepted that protein folding can occur efficiently only if a thin layer of water is immobilized, or structured, around the protein surface, the relaxation time of such a layer being higher than the bulk, however it is not clear how the dielectric properties of water can change depending on the size of the layer [40]. Taking into account two different structures in liquid water may help in facing this complex issue.

We have also faced the problem of the difficulty of fitting both the real and the imaginary part of permittivity with the same parameters by adding a term to $\varepsilon''(\omega)$ proportional to the frequency. The need for such a term opens a very intriguing issue about its physical meaning. It turns out that, in order to fit the data a term has to be added to the dissipation even for dielectrics, this term is different from the contribution due to the "free" charges used for conductors. The discussion of such an important issue is out of the scope of this paper, but it can shed a light on the nature of the organized phase of liquid water.

The picture proposed in this paper also allows fitting the data of permittivity of chlorides solutions in a very extended range of concentrations. Our results are fully compatible with the results shown in the literature showing that in electrolytic solutions, except for the case of LiCl, the dispersion and absorption are higher than in pure water. LiCl, on the opposite, is more transparent than pure water to THz radiation. We motivate these results with the lower disordering effect of LiCl on the cluster of liquid water as can be seen in Fig.2. Explanations different from the existence of two populations and based on claimed modification of the hydration spheres are, in fact, useless when the molarity exceeds 1M because of the very complex electrostatic picture of the system and the ions-recombination effect. We introduced the fraction of structured molecules of solvent N, which includes the very complex electrostatic description of concentrated solutions.

The proposed picture, strongly supports the existence of two components in liquid water.

## Appendix A

We used ATR-FTIR spectroscopy to obtain vibrational spectra of pure water and of chloride solutions of monovalent ions Li+, Na+, K+, Cs+. To prepare the solutions in bi-distilled water the salts have been purchased by Sigma-Aldrich. We used a Shimadzu-IR Affinity-1S equipped with a single reflection ATR diamond cell. Measurements were made at room temperature. For each spectrum 45 scans were recorded at a resolution of 4 cm$^{-1}$. The preliminary reduction of the IR spectral data (ATR penetration depth compensation, baseline subtraction, normalisation,) and the fitting were performed using the Labsolution Shimadzu spectral analysis software, such as the bands deconvolution routine. The broad absorption peak in the range 2800-3800 cm$^{-1}$ can be seen as the superposition of a minimum number of sub-bands that allows a univocal deconvolution of the peak. Each sub-band has been attributed to $H_2O$ molecules having a different degree of coordination among them. In particular, the sub-band centred around 3200 cm$^{-1}$ represents the lowest energy OH intermolecular bond vibrations of the strongly correlated molecular clusters. Its area is about 48% of the whole peak for pure water at 25 °C. The intermediate and the high energy bands can be attributed to weakly bounded or independent molecules, however, the distinction between them is not yet completely clear. Moreover, the area of the highest energy peak is only about 6% of the whole. Since the area of a sub-band is proportional to the number of chromophores absorbing in that region, we assumed, in this study, to consider the normalised area of the 3200 cm$^{-1}$ peak proportional to the number N of water molecules arranged in cluster while the remaining (1-N) are associated to the remaining area.

It must be underlined that the deconvolution in three Gaussian·Lorentzian peak sub-bands is univocal because: a) three is the minimum number of sub-bands and b) it is physically meaningful. The second derivative of the spectrum between 2800 and 3800 cm$^{-1}$ shows, in fact, three minimum corresponding to the sub-structures of the spectrum (Ref. 14 and 25).

From such an analysis we obtained the values of N reported in Table I.

We also looked for confirmation of that with the multivariate curve resolution (MCR) analysis. Traditionally, MCR was conceived for evolutionary analytical data coming from a process, in our case we applied MCR on FTIR spectra of evolutionary character in the concentration direction, i.e. we considered the spectra of the solution on varying the concentration. We obtained a clear indication that the spectra can be decomposed in three contributions each of them showing a different trend with the concentration. In case of pure water we considered the evolutionary character of the spectra in temperature. In figure A.1 and A.2 are compared the results of the second derivative on the pure water spectrum at 25 °C and the MCR analysis on pure water in the range 30-60 °C. Even though it is mathematically possible to do the deconvolution of the spectrum in more than three sub-bands, as reported by many authors, the MCR analysis shows that more than three contributions lead to an un-physical representation of the spectra. The behavior of the three components as a function of the temperature deserves a separate discussion that will be the subject of a future publication.

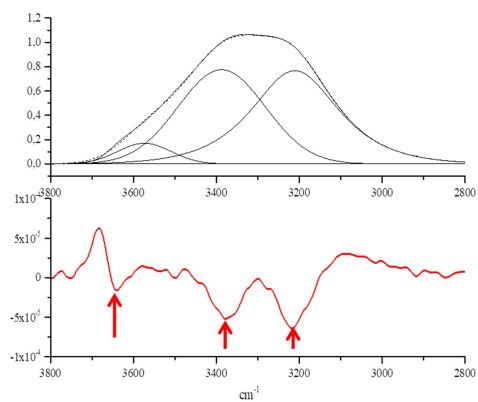

Figure A1. OH stretching band of pure water with its deconvolution sub-bands (upper panel). Three minima are clearly visible in the second derivative (lower panel) (the Savitzy-Golay smoothing slightly shift the minimum in the bottom panel with respect to the position of the sub-bands maxima).

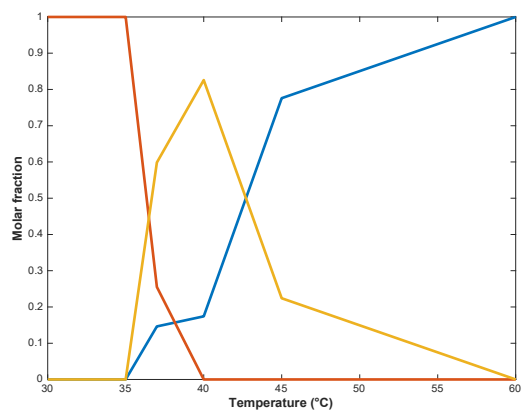

Figure A2. MCR analysis of pure water spectra vs. temperature. The red line refers to the lowest energy component, the yellow line to the intermediate one and the blue line to the high energy component. The lowest energy components have their maximum value (here normalised to 1) at low temperature and then decreases in favour of the higher energy components

**Appendix B**

In order to fit the real part of the dielectric permittivity of both pure water and chloride solutions we have used the equation:

$$\varepsilon(\omega)' = N\left[\varepsilon_{\infty 1} + \frac{(\varepsilon_{01} - \varepsilon_{\infty 1})}{1 + (\omega\tau_1)^2}\right] + (1-N)\left[\varepsilon_{\infty 2} + \frac{(\varepsilon_{02} - \varepsilon_{\infty 2})}{1 + (\omega\tau_2)^2}\right]$$

with the following constrains:

1. $\frac{\tau_1}{\tau_2} = \frac{\varepsilon_{01}}{\varepsilon_{02}} = R$
2. $\varepsilon_{\infty 2} = n^2$ being $n$ the refractive index of the specific solutions at optical frequencies.

In the case of pure water we have fitted a set of data obtained from merging our THz data with the MW data obtained by Kaatze[31]. All fit parameters were let free apart from $n^2$ which was set to values found in the literature[36,37,41], and N, whose value we obtained from the analysis of ATR-FTIR spectra (details in section 3). Since we could not find MW data for the ionic solutions to integrate with our THz measurements, when fitting their spectra we needed to add two more constrains:

3. $\tau_2^{ion} = \tau_2^{H2O} = 0.31 \, ps$
4. Being the measured static permittivity the average of the static constants of the two fractions, we can express it as: $\varepsilon_{0meas} = N\varepsilon_{01} + (1-N)\varepsilon_{02}$ and:

$$\varepsilon_{01} = \frac{R}{N(R-1)+1}\varepsilon_{0meas} = f_1$$

$$\varepsilon_{02} = \frac{1}{N(R-1)+1}\varepsilon_{0meas} = f_2$$

With a simple algebra, we get:

$$\varepsilon'(\omega) = N\left(\varepsilon_{\infty 1} + \frac{f_1 - \varepsilon_{\infty 1}}{1 + (\omega R\tau_2)^2}\right) + (1-N)\left(n^2 + \frac{f_2 - n^2}{1 + (\omega\tau_2)^2}\right)$$

with $\varepsilon_{\infty 1}$ and R as the only parameter of the fit, being the other set equal to measured values.

The same procedure on the imaginary part leads to:

$$\varepsilon''(\omega) = N\frac{f_1 - \varepsilon_{\infty 1}}{1 + (\omega R\tau_2)^2}\omega R\tau_2 + (1-N)\frac{f_1 - n^2}{1 + (\omega\tau_2)^2}\omega\tau_2 + \frac{\sigma}{\omega\varepsilon} + \xi\omega$$

Apart from the term accounting for the contribution of the ionic conductivity to the dielectric losses $\sigma/\omega\varepsilon$, where σ is static conductivity and ε the permittivity of vacuum, we had to add the linear term $\xi\omega$ which is always necessary to obtain a good fit of the data. In order to have a good match between the values obtained by both the fit of the real part and of the imaginary part, it is required to set, in $\varepsilon''(\omega)$ the same value of $\varepsilon_{\infty 1}$ as obtained from the fit of $\varepsilon'(\omega)$.

The numerical results of the fit are shown in the following tables.

| LiCl | ε' | ε" | ε' | ε" | ε' | ε" | ε' | ε" |
|---|---|---|---|---|---|---|---|---|
| | 1M | | 2M | | 3M | | 4M | |
| $\varepsilon_{0exp}$ | 64.2 ± 0 | | 53.23 ± 0 | | 45.3 ± 0 | | 38.64 ± 0 | |
| $\varepsilon_{01}$ | 130.2 ± 0.5 | 130.2 ± 0 | 109.9 ± 0.1 | 109.9 ± 0 | 95.7 ± 0.2 | | 83.1 ± 0.5 | 83.1 ± 0 |
| $\tau_1$ | 7.2 ± 0.3 | 7.2 ± 0.2 | 6.1 ± 0.2 | 6.4 ± 0.2 | 5.3 ± 0.2 | | 4.5 ± 0.2 | 5.3 ± 0.2 |
| $\varepsilon_{\infty 1}$ | 6.49 ± 0.01 | 6.49 ± 0 | 6.75 ± 0.01 | 6.75 ± 0 | 7.04 ± 0.01 | | 7.54 ± 0.01 | 7.54 ± 0 |
| R | 23.03 ± 0.09 | 23.15 ± 0.03 | 19,40± 0.07 | 20.47 ± 0.02 | 16.87 ± 0.05 | | 14,31 ±0.05 | 16.90 ± 0.04 |
| $\varepsilon_{02}$ | 5.65 ± 0.05 | 5.63 ± 0.03 | 5.67 ± 0.03 | 5.36 ± 0.01 | 5.67 ± 0.01 | | 5.80 ± 0.01 | 4.91 ± 0.01 |
| $\tau 2$ | 0.312 ± 0 | 0.312 ± 0 | 0.312 ± 0 | 0.312 ± 0 | 0.312 ± 0 | | 0.312 ± 0 | 0.312 ± 0 |
| $\varepsilon_{\infty 2}$ | 1,8 ± 0 | 1,8 ± 0 | 1,82 ± 0 | 1,82 ± 0 | 1,842 ± 0 | | 1.864 ± 0 | 1.864 ± 0 |
| N | 0,47 ± 0 | 0,47 ± 0 | 0,456 ± 0 | 0,456 ± 0 | 0,44 ± 0 | | 0.425 ± 0 | 0.425 ± 0 |
| χ | - | 0.409 ±0.006 | - | 0.504 ± 0.005 | - | | - | 0.698±0.009 |
| $S_1$ | 123.7 ± 0.5 | 123.8 ± 0.5 | 103.2 ± 0.1 | 103.1 ± 0.1 | 88.7 ± 0.2 | | 75.5 ± 0.4 | 75.5 ± 0.5 |
| $S_2$ | 3.84 ± 0.06 | 4.05 ± 0.06 | 3.85 ± 0.04 | 3.5 ± 0.1 | 3.83 ± 0.03 | | 3,941 ±0.007 | 3.051 ± 0.005 |
| σ | - | 7.86 ± 0 | - | 12.34 ± 0 | - | * | - | 17.85 ± 0 |

| LiCl | ε' | ε" | ε' | ε" | ε' | ε" |
|---|---|---|---|---|---|---|
| | 7M | | 10M | | 12M | |
| $\varepsilon_{0exp}$ | 26.6 ± 0 | | 23 ± 0 | | 22 ± 0 | |
| $\varepsilon_{01}$ | 59 ± 1 | 59 ± 0 | 51.8 ± 1.2 | | 49.7 ± 1.2 | 49.7 ± 0 |
| $\tau_1$ | 3.2 ± 0.1 | 3.7 ± 0.1 | 2.65 ± 0.09 | | 2.49 ± 0.09 | 3.4 ± 0.1 |
| $\varepsilon_{\infty 1}$ | 8.63 ± 0.01 | 8.63 ± 0 | 9.71 ± 0.02 | | 10.61± 0.02 | 10.61 ± 0 |
| R | 10.22 ± 0.03 | 11.83 ± 0.03 | 8.47± 0.03 | | 7.97 ± 0.03 | 10.99 ± 0.08 |
| $\varepsilon_{02}$ | 5.78 ± 0.08 | 5.00 ± 0.07 | 6.1 ± 0.1 | | 6.2 ± 0.1 | 4.52 ± 0.07 |
| $\tau 2$ | 0.312 ± 0 | 0.312 ± 0 | 0.312 ± 0 | | 0.312 ± 0 | 0.312 ± 0 |
| $\varepsilon_{\infty 2}$ | 1.926 ± 0 | 1.926 ± 0 | 1.988 ± 0 | | 2.03 ± 0 | 2.03 ± 0 |
| N | 0.39 ± 0 | 0.39 ± 0 | 0.37 ± 0 | | 0.363 ± 0 | 0.363 ± 0 |
| χ | - | 0.78 ±0.01 | - | | - | 1.19± 0.03 |
| $S_1$ | 50 ± 1 | 50.5 ± 0.9 | 42.1 ± 1.2 | | 39.0 ± 1.2 | 38.7 ± 1.1 |
| $S_2$ | 3.86 ± 0.06 | 3.07 ± 0.05 | 4.12 ± 0.09 | | 4.2 ± 0.1 | 2.49 ± 0.05 |
| σ | - | 17.1 ± 0 | - | * | - | 11.5 ± 0 |

*No data available

| NaCl | Eps' | Eps" | Eps' | Eps" | Eps' | Eps" | Eps' | Eps" | Eps' | Eps" |
|---|---|---|---|---|---|---|---|---|---|---|
| | 1M | | 2M | | 3M | | 4M | | 5M | |
| $\varepsilon_{0exp}$ | 63.48 ± 0 | | 53.67 ± 0 | | 45.39 ± 0 | | 39.2 ± 0 | | 34.5 ± 0 | |
| $\varepsilon_{01}$ | 133.8 ± 0.6 | 133.8 ± 0 | 122.2 ± 0.4 | 122.2 ± 0 | 106.50 ± 0.05 | 106.5 ± 0 | 93.4 ± 0.3 | 93.4 ± 0 | 82.1 ± 0.5 | 82.1 ± 0 |
| $\tau_1$ | 7.1 ± 0.3 | 6.6 ± 0.2 | 6.3 ± 0.2 | 5.8 ± 0.2 | 5.3 ± 0.2 | 5.15 ± 0.17 | 4.6 ± 0.2 | 4.6 ± 0.1 | 3.9 ± 0.1 | 4.1 ± 0.01 |
| $\varepsilon_{\infty 1}$ | 6.77 ± 0.02 | 6.77 ± 0 | 7.51± 0.02 | 7.51 ± 0 | 8.22 ± 0.02 | 8.22± 0 | 9.95 ± 0.02 | 9.95 ± 0 | 9.62± 0.02 | 9.62± 0 |
| R | 22.6 ± 0.1 | 21.05± 0.04 | 20.2 ± 0.1 | 18.76 ± 0.04 | 16.85 ± 0.07 | 16.51 ± 0.03 | 14.66 ± 0.07 | 14.73 ± 0.02 | 12.61 ± 0.05 | 13.22 ± 0.02 |
| $\varepsilon_{02}$ | 5.92 ± 0.06 | 6.05 ± 0.04 | 6.05 ± 0.05 | 6.51 ± 0.03 | 6.32 ± 0.03 | 6.45 ± 0.01 | 6.37 ± 0.01 | 6.34 ± 0.01 | 6.51 ± 0.02 | 6.2 ± 0.03 |
| $\tau 2$ | 0,312 ± 0 | 0,312 ± 0 | 0,312 ± 0 | 0,312 ± 0 | 0,312 ± 0 | 0,312 ± 0 | 0,312 ± 0 | 0,312 ± 0 | 0,312 ± 0 | 0,312 ± 0 |
| $\varepsilon_{\infty 2}$ | 1,8 ± 0 | 1,8 ± 0 | 1,83 ± 0 | 1,83 ± 0 | 1,863 ± 0 | 1,863 ± 0 | 1,893 ± 0 | 1,923 ± 0 | 1,923 ± 0 | 1,923 ± 0 |
| N | 0,45 ± 0 | 0,45 ± 0 | 0,41 ± 0 | 0,41 ± 0 | 0,39 ± 0 | 0,39 ± 0 | 0,377 ± 0 | 0,377 ± 0 | 0,37 ± 0 | 0,37 ± 0 |
| $\chi$ | - | 0.286 ±0.009 | - | 0.36 ± 0.01 | - | 0.516 ± 0.009 | - | 0.658± 0.008 | - | 0.841 ± 0.009 |
| $S_1$ | 127.1 ± 0.6 | 127.1 ± 0.6 | 114.7 ± 0.4 | 114.7 ± 0.4 | 98.28 ± 0.07 | 98.29 ± 0.03 | 84.5 ± 0.2 | 84.5 ± 0.03 | 72.5 ± 0.5 | 72.5± 0.5 |
| $S_2$ | 4.12 ± 0.08 | 4.55 ± 0.06 | 4.22 ± 0.07 | 4.68 ± 0.05 | 4.45 ± 0.05 | 4.46 ± 0.03 | 4.48 ± 0.03 | 4.451 ± 0.007 | 4.592± 0.002 | 4.29± 0.01 |
| $\sigma$ | - | 7.81 ± 0 | - | 13.29 ± 0 | - | 17.63 ± 0 | - | 20.63 ± 0 | - | 22.47 ± 0 |

| KCl | Eps' | Eps" | Eps' | Eps" | Eps' | Eps" | Eps' | Eps" |
|---|---|---|---|---|---|---|---|---|
| | 1M | | 2M | | 3M | | 4M | |
| $\varepsilon_{0exp}$ | 67.2 ± 0 | | 57.7 ± 0 | | 51.16 ± 0 | | 44.8 ± 0 | |
| $\varepsilon_{01}$ | 139.0 ± 0.7 | 139 ± 0 | 127.5 ± 0.5 | 127.5 ± 0 | 120.6 ± 0.4 | | 109.3 ± 0.1 | 109.3 ± 0 |
| $\tau_1$ | 7.2 ± 0.3 | 7.2 ± 0.3 | 6.3 ± 0.2 | 6.2 ± 0.2 | 5.5 ± 0.2 | | 4.9 ± 0.2 | 4.9 ± 0.2 |
| $\varepsilon_{\infty 1}$ | 6.34 ± 0.01 | 6.34 ± 0 | 7.16 ± 0.02 | 7.16 ± 0 | 7.78 ± 0.03 | | 8.36 ± 0.03 | 8.36 ± 0 |
| R | 23.1 ± 0.1 | 22.96± 0.03 | 20.17 ± 0.08 | 19.94± 0.03 | 17.8 ± 0.1 | | 15.80± 0.08 | 15.68 ± 0.03 |
| $\varepsilon_{02}$ | 6.02 ± 0.06 | 6.0 ± 0.1 | 6.32 ± 0.05 | 6.39 ± 0.03 | 6.78 ± 0.06 | | 6.91 ± 0.04 | 6.95 ± 0.02 |
| $\tau 2$ | 0.312 ± 0 | 0.312 ± 0 | 0.312 ± 0 | 0.312 ± 0 | 0.312 ± 0 | | 0,312 ± 0 | 0,312 ± 0 |
| $\varepsilon_{\infty 2}$ | 1,8 ± 0 | 1,8 ± 0 | 1,83 ± 0 | 1,83 ± 0 | 1,877 ± 0 | * | 1,91 ± 0 | 1,91 ± 0 |
| N | 0,46 ± 0 | 0,46 ± 0 | 0,424 ± 0 | 0,424 ± 0 | 0,39 ± 0 | | 0,37± 0 | 0,37± 0 |
| $\chi$ | - | 0.407 ± 0.008 | - | 0.518 ± 0.008 | - | | - | 0.76± 0.01 |
| $S_1$ | 132.6 ± 0.7 | 132.6 ± 0.8 | 120.3 ± 0.5 | 120.3 ± 0.5 | 112.8 ± 0.4 | | 100.9 ± 0.1 | 100.67 ± 0.08 |
| $S_2$ | 4.22 ± 0.06 | 4.3 ± 0.1 | 4.49 ± 0.07 | 4.56 ± 0.03 | 4.90 ± 0.08 | | 5.00 ± 0.06 | 5.04 ± 0.02 |
| $\sigma$ | - | 11.27 ± 0 | - | 21.7 ± 0 | - | | - | 37.45 ± 0 |

*No data available

| CsCl | Eps' | Eps" | Eps' | Eps" | Eps' | Eps" | Eps' | Eps" |
|---|---|---|---|---|---|---|---|---|
| | 1M | | 2M | | 3M | | 6M | |
| $\varepsilon_{0exp}$ | 69.1 ± 0 | | 60.62 ± 0 | | 52.88 ± 0 | | 41 ± 0 | |
| $\varepsilon_{01}$ | 142.1 ± 0.8 | 142.1 ± 0 | 131.1 ± 0.5 | 131.1 ± 0 | 118.1 ± 0.3 | 118.1 ± 0 | 105.64 ± 0.06 | |
| $\tau_1$ | 6.7 ± 0.2 | 6.8 ± 0.2 | 5.8 ± 0.2 | 5,8 ± 0.2 | 4.7 ± 0.2 | 5.7 ± 0.2 | 3.9 ± 0.2 | |
| $\varepsilon_{\infty 1}$ | 5.84 ± 0.01 | 5.84 ± 0 | 5.83 ± 0.01 | 5.83 ± 0 | 5.56 ± 0.02 | 5.56 ± 0 | 7.44 ± 0.08 | |
| R | 21.37 ± 0.08 | 21.93 ± 0.02 | 18.71 ± 0.05 | 19.76 ± 0.02 | 15.19 ± 0.06 | 18.39 ± 0.07 | 12.5 ± 0.1 | |
| $\varepsilon_{02}$ | 6.65 ± 0.06 | 6.48 ± 0.04 | 7.00 ± 0.05 | 6.66 ± 0.03 | 7.77 ± 0.05 | 6.43 ± 0.04 | 8.44 ± 0.07 | |
| $\tau_2$ | 0.312 ± 0 | 0.312 ± 0 | 0.312 ± 0 | 0.312 ± 0 | 0.312 ± 0 | 0.312 ± 0 | 0.312 ± 0 | |
| $\varepsilon_{\infty 2}$ | 1,81 ± 0 | 1,81 ± 0 | 1,842 ± 0 | 1,842 ± 0 | 1,873 ± 0 | 1,873 ± 0 | 1,966 ± 0 | |
| N | 0,461 ± 0 | 0,461 ± 0 | 0,432 ± 0 | 0,432 ± 0 | 0,409 ± 0 | 0,409 ± 0 | 0,335 ± 0 | |
| χ | - | 0.299 ± 0.006 | - | 0.367 ± 0.006 | - | 0.69 ± 0.02 | - | |
| $S_1$ | 136.3 ± 0.8 | 136.3 ± 0.8 | 125.3 ± 0.6 | 125.3 ± 0.6 | 112.5 ± 0.3 | 112.5 ± 0.4 | 98.2 ± 0.1 | |
| $S_2$ | 4,84 ± 0.08 | 4.67 ± 0.06 | 5.16 ± 0.07 | 4.82 ± 0.05 | 5.9 ± 0.07 | 4.56 ± 0.06 | 6.47 ± 0.09 | |
| σ | x | 11.8 ± 0 | x | 19.5 ± 0 | x | 36.5 ± 0 | x | * |

*No data available

**Acknowledgments.** We thank F. Marini for his valuable help in MCR analysis and Islam MD Deen for help in THz-TDS measurements.